%% file: main.tex
\documentclass[aps,reprint,prx,longbibliography]{revtex4-1}
\pdfoutput=1
\usepackage[utf8]{inputenc}
\usepackage{amsmath,amssymb,mathrsfs,amsthm}
\usepackage{bbm}
\usepackage{chemarrow}
\usepackage{graphicx}
\usepackage{accents}
\usepackage{sidecap}
\usepackage[colorlinks=true,allcolors=blue]{hyperref}
\usepackage{breakurl}
\usepackage{soul}
\usepackage{float}
\usepackage[shortlabels]{enumitem}
\usepackage{epstopdf}

\input{macros}

\begin{document}

\title{Emergence of Newtonian Deterministic Causality from Stochastic\\ Motions in Continuous Space and Time}

\author{Bing Miao}
\affiliation{Center of Materials Science and Optoelectronics Engineering, College of Materials Science and Opto-Electronic Technology, University of Chinese Academy of Sciences, Beijing 100049, P.R.C.}

\author{Hong Qian}
\affiliation{Department of Applied Mathematics, University of Washington, Seattle, WA 98195, U.S.A.}

\author{Yong-Shi Wu}
\affiliation{Department of Physics \& Astronomy, University of Utah, Salt Lake City, UT 84112, U.S.A.}


\begin{abstract}
Since Newton's time, deterministic causality has been considered a crucial prerequisite in any fundamental theory in physics. In contrast, the present work investigates stochastic dynamical models for motion in one spatial dimension, in which Newtonian mechanics becomes an emergent property: We present a coherent theory in which a Hamilton-Jacobi equation (HJE) emerges in a description of the evolution of entropy $-\phi(x,t)=\epsilon \log$(Probability) of a system under observation and in the limit of large information extent $\epsilon^{-1}$ in homogeneous space and time. The variable $\phi$ represents a non-random high-order statistical concept that is distinct from probability itself as $\epsilon=0$; the HJE embodies an emergent law of deterministic causality in continuous space and time with an Imaginary Scale symmetry $(t,x,\phi)\leftrightarrow (it,ix,-i\phi)$. $\phi(x,t)$ exhibits a nonlinear wave phenomenon with a mathematical singularity in finite time, overcoming which we introduce viscosity $\epsilon(\partial^2\phi/\partial x^2)$ and wave $i\epsilon(\partial^2 \phi/\partial x^2)$ perturbations, articulating dissipation and conservation, which break the Imaginary Scale symmetry: They lead to the Brownian motion and Schr\"{o}dinger's equation of motion, respectively.  Last but not least, Lagrange's action in classical mechanics acquires an entropic interpretation and Hamilton's principle is established.

\end{abstract}

\maketitle


\section{Introduction}

Emergent phenomenon defined by a limit theorem in a class of mathematical models usually exhibits certain law-like behavior or property, that is not true before the limit is reached \cite{yanglee,leeyang,anderson72,Chibbaro-book,qatw}.  An obvious example is that condition $N,V\to\infty$ with fixed $\rho=N/V$ defines the thermodynamic limit, where $N$ is the particle number and $V$ is the volume of a mechanical system.  In quantum statistics the Kubo-Martin-Schwinger (KMS) thermodynamic Green's function is a non-trivial case \cite{kubo_1957, martin_schwinger}.

The profound insights from statistical physics, particularly in the theory of phase transition, have provided us with an intriguing recipe for generating new principles in the age of machine learning via the idealization of data {\em ad infinitum} \cite{yangqian_22,yangqianAP}.  In the traditional statistical thermodynamics, the ``infinity'' has always been identified with the largeness of a macroscopic system, whose equilibrium possesses an Eulerian homogeneous degree-$1$ entropy function $S(E,V,N)$ \cite{callen-book}.  In a recent series of papers on the thermodynamics of small systems \cite{qian_jctc,lu-qian-22,qian_holographic,qian-21}, it was suggested that T. L. Hill's theory \cite{Hill_SST_Book,bedeaux-book} should be framed as for repeated measurements in time, with ``infinity'' being identified with a scientific measurement over a $\Delta t$ which, no matter how small, nevertheless is very large in a continuous time that is infinitely divisible.

A novel thermodynamic function $\mathscr{E}$, called {\em subdivision potential}, figures prominently in Hill's theory: It is the Legendre-Fenchel (LF) transform of a non-homogeneous entropy function $S(E,V,N)$:
\begin{subequations}
\label{equation1}
\begin{equation}
   \mathscr{E}(\beta,p,\mu) =
   \frac{1}{\beta} \inf_{E,V,N}
   \Big\{\beta \big(E+ pV -\mu N\big) -S(E,V,N) \Big\}.
\end{equation}
It is easy to verify that Eq. (\ref{equation1}a) in terms of the infimum replaces the standard Legendre transform based on derivatives in physics \cite{zia_paper}.  This is a mathematical refinement from modern convex analysis \cite{rockafellar-book} according to which when $S(\lambda E,\lambda V,\lambda N)=\lambda S(E,V,N)$ for all $\lambda$,
\begin{equation}
  \mathscr{E}(\beta,p,\mu)= \left\{ \begin{array}{ccc}
   0  && \mu=f(\beta,p) \\
   -\infty && \text{elsewhere}
   \end{array} \right.
\end{equation}
where $\mu=f(\beta,p)$ is known as a thermodynamic {\em equation of state}, the LF dual to the homogeneous entropy $S(E,V,N)$:\footnote{To show Eq. \ref{equation1}c, we first note that $\mu=f(\beta,p):=N^{-1} \inf_{E,V}\{E+pV-\beta^{-1}S(E,V,N)\}$. Then $\inf_{\beta,p}\big\{\beta\big( E+pV-Nf(\beta,p)\big)\big\}$ $=\inf_{\beta,p}\big\{\beta E+\beta pV-\inf_{E',V'}\{\beta E'+\beta pV'-S(E',V',N)\}\big\}$ $= \sup_{E',V'}\big\{ \inf_{\beta,p}\{ \beta (E-E')+\beta p(V-V')+S(E',V',N)\} \big\}$ $=S(E,V,N)$. This completes the LF duality between the functions $S=S(E,V,N)$ and $\mu=f(\beta,p)$, even though they look very different.
}
\begin{equation}
     S(E,V,N) = \inf_{\beta,p,\mu}
      \Big\{\beta \big(E+ pV -\mu N\big)
      \Big| \mu-f(\beta,p)=0 \Big\}.
\end{equation}
\end{subequations}

We consider the limit of random motions in continuous space and time in the present work.  The authors are aware of the fact that in mathematical literature, the notion of random processes in continuous space and time has been rigorously defined in a much more refined way via Brownian motions \cite{wang-uhlenbeck} and L\'{e}vy processes with their Le\'{v}y-Khinchin characteristic exponents \cite{schilling-book}.  From the perspective of a complex world at $10^{-100}$ meter and $10^{-100}$ second whose existence we postulate, any scientific $\Delta x$ and $\Delta t$ are extensive quantities whose ratio, $\Delta x/\Delta t$, defines a ``macroscopic'' velocity.  We seek emergent thermodynamic law(s) in this mathematical idealization and find a Hamilton-Jacobi equation (HJE) for entropy, shown in Eq. (\ref{gHJE}), which arises in this context.  In classical physics, HJE is an integral part of Newtonian mechanics \cite{landau_mechanics, evans-book} that obeys strictly the law of causality with determinism. The success of Newtonian mechanics has made many people to believe in their subconsciousness that the Principle of Deterministic Causality (PDC) should be a built-in fundamental in any physics theory \cite{Max_Born_book, J_Monod_book,David_Ruelle_book, Walter_Hehl_book}. The development of the special theory of relativity seemed to have strengthened this belief \cite{debroglie-book}.

In Hamilton's canonical formulation of Newton's mechanics, there are two distinct but equivalent ways to describe the time evolution of a mechanical system. One is by Hamilton's equations, which are $2n$ first-order ordinary differential equations (ODEs) for canonical variables, generalized coordinates $\vq=(q_1,\cdots,q_n)$ and momenta $\vp=(p_1,\cdots,p_n)$ with $n$ being the number of degrees of freedom. This is equivalent to Newton's equations of motion with $n$ second-order ODEs, one for each coordinate. Another way is the HJE, which is a first-order partial differential equation (PDE) for the action function $S(\vq,t)$, not to be confused with the action functional that is an integral, whose $(n+1)$ independent variables are coordinates $\vq$ and time $t$.  It is of the form ${\partial S}/{\partial t} + H(\vq,\partial S/\partial\vq) =0$, where $H(\vq,\vp)$ is the Hamiltonian function. The definition of the action function $S$ is the following \cite{landau_mechanics}: Consider open paths in configuration space with the initial point $\vq(0)$ fixed; each path satisfies the equations of motion. Then the value of the integral along a path,
\begin{subequations}
\label{equation2}
\begin{eqnarray}
    &\displaystyle
    S\big(\vq,t\big)= S(\vq,0)+ \int_0^t L\big(\vq(s),\dot{\vq}(s)\big)
     \rd s, \text{ in which } \hspace{.3in}
\\
    &\displaystyle
     L\big(\vq,\dot{\vq}\big) = \sup_{\vp} \Big\{
       \dot{\vq}\cdot\vp - H(\vq,\vp)\Big\}, \
       \dot{\vq} = \frac{\partial H}{\partial\vp},
\\
    &\displaystyle
    H\big(\vq,\vp\big) = \sup_{\dot{\vq}} \Big\{
       \dot{\vq}\cdot\vp - L(\vq,\dot{\vq})\Big\}, \
       \vp = \frac{\partial L}{\partial\dot{\vq}},
\end{eqnarray}
\end{subequations}
is only a function of the final point $\vq$ and final time $t$. An open path has a $\dot{\vq}(0)$, which is one-to-one related to $\partial S(\vq,0)/\partial\vq=\vp(0)=\partial L(\vq(0),\dot{\vq}(0))/\partial\dot{\vq}$ for convex $H(\cdot,\vp)$ and $L(\cdot,\dot{\vq})$ via LF duality in (\ref{equation2}b), (\ref{equation2}c).  This means that dynamics represented by an HJE either starts as a geometric point with $\vq(0)$ and $\dot{\vq}(0)$, or from an $S(\vq,0)$ with finite support.  HJE embodies the insight of Huygens' principle that unifies geometric and wave optics; this will be the motivation for the perturbation treatment in Sec. \ref{sec:III}.

A well-known theorem in variational calculus asserts that if a particular type of solution, called a complete integral of HJE \footnote{Solutions to a first-order PDE depends on an arbitrary function $S(\vq,0)$. The {\em complete integral} to an HJE is a ``general'' solution in the form of $S(\vq,t,\alpha_1,\cdots,\alpha_n)$ in which the $n$ arbitrary constants are related to the arbitrary function.  On the other hand, {\em a characteristic initial condition} is a particular $S(\vq,0)$ with $H\big(\vq,\partial S(\vq,0)/\partial\vq\big)=E$, a constant independent of $\vq$.  Then $S(\vq,t)=S(\vq,0)-Et$ \cite{evans-book}. }, is known then one can reconstruct the solutions of Hamilton's canonical equations, thus obtaining Newtonian motion. In this way, we see that the HJE actually embodies the PDC. Nobody thinks that PDC is satisfied in stochastic dynamical systems.  The present paper suggests that PDC could be an emergent property from random motions in continuous space and time as a mathematical limit, {\em \`{a} la} Anderson \cite{anderson72}.  See Appendix \ref{app-B} for a very simply illustrative example.

\section{A Hamilton-Jacobi equation under translational symmetry with anisotropy}

For a simple biased random walk with spatial and temporal resolutions $\Delta x$ and $\Delta t$ and probabilities $p$, $q$, and $r=1-p-q$ for stepping forward, backward, and not moving, the law of total probability dictates the probability distribution $P_x(t)$ to follow the time stepping:
\[
     P_x(t+\Delta t) = p P_{x-\Delta x}(t) + r P_x(t) + q P_{x+\Delta x}(t).
\]
We adopt translational symmetry so that $p$ and $q$ are constants independent of $x$; $p\neq q$ however represents an anisotropy \footnote{In one spatial dimension, the non-zero $p\neq q$ can always be eliminated via a translation.  This is equivalent to redefining $y'=y+V/(2D)$ as the conjugate momentum in Eq. (\ref{HJE}). However, in a spatial dimension greater than one, a Lorentz-type rotational motion cannot be eliminated with such a transformation; this corresponds to stochastic motion without detailed balance \cite{ge_qian_ijmpb}.  For this reason, we have kept the first-order term in our analysis.} In the continuous space and time limit of $\Delta x$, $\Delta t\to 0$ with $\Delta x/\Delta t$ fixed, it can be found in all standard textbooks that a transport equation $\partial P_x(t)/\partial t = -V\partial P_x(t)/\partial x$ emerges, where $V=(p-q)\Delta x/\Delta t$. Smoluchowski-Einstein diffusion arises only when $p=q$ under the supposition that $(1-r)(\Delta x)^2/\Delta t=2D$ is fixed: the $\Delta t\to 0$ is much faster than $\Delta x\to 0$ in this case.

There is another more fundamental treatment that encompasses much more: Let
$P_x(t)\sim e^{-\phi(x,t)/\epsilon}$ where $\epsilon^{-1}$ represents the number of measurements in time on $x$ that tends to infinity, and correspondingly $\phi(x,t)$ as a {\em neg-entropy evolution} satisfies
(see Appendix \ref{app-C} for details)
\begin{eqnarray}
 \frac{\partial\phi(x,t)}{\partial t}
   &+& \gamma\log\left\{1 +
p\Big(e^{\eta(\partial\phi/\partial x)}
   -1\Big) \right.
\nonumber\\
   &+& \left. q\Big(e^{-\eta(\partial\phi/\partial x)}
    -1\Big)\right\} = 0,
\label{gHJE}
\end{eqnarray}
where $\gamma=\epsilon/\Delta t$ and $\eta=\Delta x/\epsilon$. In other words, with $\epsilon^{-1}$ representing the information extent, we consider the limit of all $\Delta x, \Delta t$ and $\epsilon\to 0$ with $\gamma$ and $\eta$ fixed.  The emergence and existence of $\phi$ is a highly non-trivial mathematical matter \cite{dembo-book}.

If one further asserts that in the limit of $\Delta t$, $\Delta x\to 0$, the motion $x(t)$ is a differentiable function of $t$, then Eq. (\ref{gHJE}) becomes $\partial\phi(x,t)/\partial t+ V\partial\phi/\partial x=0$, where $V=(p-q)\eta\gamma$ stands for classical particle motion with constant velocity.  If, however, one assumes a non-differentiable but nevertheless continuous motion $x(t)$, such as a Brownian motion, with diffusion coefficient $\epsilon D$ and a drift $V$ \cite{DurrettBook}, then
\begin{equation}
\label{HJE}
    \frac{\partial\phi(x,t)}{\partial t}
   + H\left(x,\frac{\partial\phi}{\partial x} \right) = 0, \text{ where }\,
   H(x,y) = Vy+Dy^2,
\end{equation}
which implies a Hamiltonian dynamics with $Dy^2$ and $Vy$ analogous to kinetic energy of a point mass and a velocity-dependent force.
The simple $H(x,y)$ is independent of $x$ in the case with translational symmetry.

Eq. (\ref{HJE}) implies a Newtonian second-order equation of motion, whose origin is now clear ({\em e.g.} Appendix \ref{app-B}): Langevin type motion with diffusion and drift
\begin{equation}
\label{newton}
   \rd X(t) = V\rd t + \sqrt{2\epsilon D}\, \rd W(t), \  \  X(0)=x_0,
\end{equation}
in the limit of $\epsilon\to 0$ recovers $\rd X(t)/\rd t =V$. However, if there is an observation, a measurement that constrains $X(t_1)=x_1\neq x_0+Vt_1$ where $t_1>0$, {\em i.e.}, a {\em conditioning} in the theory of probability, then the motion in the limit of $\epsilon\to 0$ becomes $\rd^2 X(t)/\rd t^2=0$, whose solutions now include $X(t)=x_0+(x_1-x_0)t/t_1$. Thermodynamics is about systems under constraints \cite{callen-book}; Newtonian motion is a consequence of stochastic dynamics (\ref{newton}) under the constraint $X(t_1)=x_1$ then taking the limit $\epsilon\to 0$ that eliminates randomness and gains PDC.

\subsection{Finite time singularity}

There are peculiar features to the solutions of Eq. (\ref{gHJE}):  First,
they possesses all the essential behavior of the convex function $\frac{(x-Vt)^2}{4Dt}$ as an explicit solution to (\ref{HJE}): At a local minimum $x^*(t)$ of $\phi(x,t)$, $\phi\big(x^*(t),t\big)$ is actually independent of $t$. The $x^*(t)$ moves with velocity $V=\gamma\eta(p-q)$. Furthermore the curvature at $x^*(t)$ satisfies a quadratic differential equation,
\begin{eqnarray}
\label{eq-4-curv}
    &\displaystyle
    \frac{\rd}{\rd t} \left(\frac{\partial^2\phi(x,t)}{\partial x^2}\right)_{x=x^*(t)} = -2D \left(\frac{\partial^2\phi}{\partial x^2}
    \right)_{x=x^*(t)}^2,
\\
    &\displaystyle
    D=\frac{1}{2}\gamma\eta^2\big(p+q-(p-q)^2 \big).
\nonumber
\end{eqnarray}
The $(\partial^2\phi/\partial x^2)$ at $x^*(t)$ blows up in finite time at \footnote{Denoting $\partial^2\phi(x^*(t),t)/\partial x^2 =w(t)$, Eq. (\ref{eq-4-curv}) is $\rd w/\rd t=-2Dw^2$, whose solution is $w(0)[1+2Dw(0)t]^{-1}$.  Therefore, if $w(0)<0$, $w(t)$ blows up at finite time $t^*=|w(0)|^{-1}/2D$. If $w(0)>0$, then $w(t)\sim (2Dt)^{-1}$ approaches $0$ extremely slow as $t\to\infty$, and it blows up at a negative time
$-t^*$. }
\begin{equation}
       |t^*|=\frac{1}{2D|\phi''_{xx}\big((x^*(0),0\big)|}.
\label{t-blow}
\end{equation}
The mathematical solution to (\ref{gHJE}) ``blows up in finite time''.   While a convex function flattens with time at its minimum, a concave function sharpens at its maximum with time and becomes singular at $t^*$.

\subsection{Dispersive wave packet}

In Eq. (\ref{gHJE}) $\rd\phi(x^*(t),t))/\rd t\equiv 0$ for all $t$, another key feature of the solution.  Noting that in relation to the probability $P_x(t)$,
\[
   P_x(t) = \mathcal{A}_{\epsilon}(t) e^{-\phi(x,t)/\epsilon},
   \text{ where } \mathcal{A}^{-1}_{\epsilon}(t) = \int_{\mathbb{R}}
     e^{-\phi(x,t)/\epsilon} \rd x.
\]
As a logarithmic probability,
\begin{equation}
\label{6}
     \epsilon \log P_x(t) = -\phi(x,t)
     +\epsilon\log\mathcal{A}_{\epsilon}(t),
\end{equation}
its ``width'' increases while its ``height'' should simultaneously decrease.  However, in the limit of $\epsilon\to 0$, these two modes of change are no longer strictly coupled; there is a separation of the time scales.  The last term in (\ref{6}) is of the order $\mathcal{O}(\epsilon\log\epsilon)$.  The motion represented by $\phi(x,t)$ is non-random; it is best described as a {\em dispersive wave packet}. Dissipation comes as a phenomenon of the next order. This is the reason we say that the motion of $\phi$ represented by HJE is distinct from the probabilistic description of random motions; it describes an emergent behavior.

In the limit of infinitely large extent as $\epsilon^{-1}\to\infty$ in continuous space and time, dissipative, random particle movements transcend into a new type of dynamics that exhibits {\em dispersive wave motion}; the cardinal rule of probability normalization is lost. Instead, the emergent dynamics under translational symmetry characterized by $\partial\phi/\partial t + H(\partial\phi/\partial x)=0$ is invariant under transformation $(t,x,\phi)\to(\lambda t,\lambda x,\lambda^{-1}\phi)$, which is related to the aforementioned Eulerian homogeneity. Even more interestingly and in anticipation of Sec. \ref{sec:III}, the Eulerian homogeneity implies that if $\phi(x,t)$ is a solution, then $-i\phi(ix,it)$ is also a solution. As far as we know, such a mathematical feature has not been formally discussed in the past in physics literature, which we designate as Imaginary Scale symmetry.

\subsection{Legendre-Fenchel duality of $\phi(x,t)$  }

Since $\phi(x,t)$ is Eulerian homogeneous degree-$1$, one immediately has the conjugate variables
\begin{equation}
   y(x,t) = \frac{\partial\phi}{\partial x} \  \text{ and } \  E(x,t)=-\frac{\partial\phi}{\partial t},
\end{equation}
satisfy an equation analogous to the Gibbs-Duhem relation in thermodynamics,
\begin{eqnarray}
    x\,\rd y(x,t) - t\,\rd E(x,t)
    &=& \big[x, \ t\big] \left[\begin{array}{cc}
     \frac{\partial^2\phi}{\partial x^2} &
     \frac{\partial^2\phi}{\partial x\partial t} \\[4pt]  \frac{\partial^2\phi}{\partial x\partial t} &
     \frac{\partial^2\phi}{\partial t^2 } \end{array}
      \right] \left[\begin{array}{c}
       \rd x \\ \rd t \end{array}\right]
\nonumber\\
   &=& \big[0, \ 0\big]\left[\begin{array}{c}
       \rd x \\ \rd t \end{array}\right] = 0.
\end{eqnarray}
One also has Fenchel-Young inequality \cite{paper-II}:
\begin{equation}
\label{eq10}
   \Xi(x,t,y,E) = \phi(x,t) + \psi(y,E)
    - xy + tE \ge 0,
\end{equation}
where
\begin{equation}
    \psi(y,E) = \sup_{x,t}
    \Big\{xy-tE-\phi(x,t)\Big\}
     =\left\{\begin{array}{ccc}
     0 &&  E=H(y)  \\
     \infty && \text{elsewhere}
    \end{array}\right..
\label{eq11}
\end{equation}
They are combined into $\phi(v,1)-vy+H(y)\ge 0$, where $v:=x/t$.  It is tempting to identify the very Hamilton-Lagrange's structure of mechanical motions as a thermodynamic ``equilibrium''; then this inequality represents a nonequilibrium concept in the thermo-doubled space of $(v,y)$ with an unadulterated arrow of time that is outside the current theoretical mechanics \cite{paper-II}.

\subsection{Stationary $\phi^{ss}(x)$ is a landscape for deterministic $x(t)$  }

In the absence of translational symmetry, in general the Hamiltonian function $H(x,y)$ that arises from stochastic dynamics in the limit of continuous space and time, where $y=\partial\phi/\partial x$, has the form \cite{ge_qian_ijmpb,Ge-Qian-2016}:
\begin{eqnarray}
   H(x,y) &=& \left(\frac{\partial H(x,0)}{\partial y}\right) y + \frac{1}{2}\left(\frac{\partial^2H(x,0)}{\partial y^2}\right) y^2
\nonumber\\
   && + \ \frac{1}{6}
   \left(\frac{\partial^3H(x,0)}{\partial y^3}\right)y^3 + \cdots
\label{Hyseries}
\end{eqnarray}
in which $H(x,0)=0$. $\dot{x}=V(x)=\partial H(x,0)/\partial y$ is the deterministic motion in the $\epsilon\to 0$ limit, which corresponds to the ``overdamped'' Hamiltonian dynamics with $y(t)\equiv 0$ \cite{ge_qian_ijmpb}.

The stationary solution to the HJE, $\varphi^{ss}(x)$, satisfies
\[
  H\left(x,\frac{\partial\phi^{ss}(x)}{\partial x}\right) = 0.
\]
Therefore from Eq. (\ref{Hyseries}) one has either $\partial\phi^{ss}(x)/\partial x=0$, or
\begin{eqnarray*}
    V(x) + \frac{1}{2}\frac{\partial^2 H(x,0)}{\partial y^2}\left(\frac{\partial\phi^{ss}}{\partial x}\right) &+& \frac{1}{6}
   \frac{\partial^3H(x,0)}{\partial y^3}\left(\frac{\partial\phi^{ss}}{\partial x}\right)^2
\\[6pt]
   &+& \cdots = 0.
\end{eqnarray*}
Thus, if $\dot{x}=V(x)$,
\begin{equation}
    \frac{\rd}{\rd t}\phi^{ss}\big(x(t)\big)
    =\left(\frac{\partial\phi^{ss}}{\partial x}\big(
     x(t)\big)\right)V(x)
     \simeq -\frac{V^2(x)}{D(x)} \le 0.
\end{equation}
The stationary $\phi^{ss}(x)$ is a ``landscape'' for the deterministic dynamics.  One can find a general proof of this result in \cite{freidlin-book,Ge-Qian-2016}.  This further substantiates the assertion that $-\phi^{ss}(x)$ is an entropy function.

\section{Regularization of HJE singularity and symmetry breaking}
\label{sec:III}

The singularity with finite-time blowup is undoubtedly nonphysical. In mathematics, an additional viscosity term with a small coefficient $\epsilon$ was introduced by Crandell and Lions \cite{lions-crandell} to ``regularize'' the singularity.

\subsection{Viscosity solution}

Since an HJE represents the limiting behavior of zero randomness of a stochastic dynamics, the regularization can be guided by recovering the dissipative, diffusion behavior in terms of a viscosity term:
\begin{equation}
\label{viscosity-soln}
      \frac{\partial\phi}{\partial t}
      = -V\left(\frac{\partial\phi}{\partial x}
      \right)-D\left(\frac{\partial\phi}{\partial x}\right)^2+\epsilon D\left(\frac{\partial^2\phi}{\partial x^2}
      \right).
\end{equation}
Introducing the Cole-Hopf transformation $f(x,t)=e^{-\phi(x,t)/\epsilon}$ and its inverse
$\phi(x,t)=-\epsilon\log f(x,t)$, we have a diffusion with drift proper:
\begin{equation}
\label{equation15}
    \frac{\partial f(x,t)}{\partial t} = -V \left(\frac{\partial f}{\partial x}\right) +\epsilon D\left(\frac{\partial^2 f}{\partial x^2}\right).
\end{equation}
Interpreting $f(x,t)\ge 0$ as a probability density function with $f(\pm\infty,t)=0$,  the solution to Eq. (\ref{equation15}) has
\[
  \int_{\mathbb{R}} f(x,t)\rd x = \text{ const for all } t.
\]

\subsection{Oscillatory wave solution}
\label{i-perturb}

It is now well-understood that stochastic dynamics without detailed balance necessarily gives rise to cyclic rotational motion \cite{zqq}.  Under a strong driving force such motion in the zero-noise limit is oscillatory \cite{cheng-qian-jsp}.  The mathematical representation of such dynamics necessarily become dynamics of ``phases'' on a torus that possesses a different topological character \cite{qw_cmp}. Unfortunately this aspect of nonequilibrium stochastic dynamics has not been as widely appreciated. Therefore, alternative to a viscosity solution, we introduce an $\epsilon$-perturbation based on the premise of an rotational motion as a fundamentally different alternative to Eq. (\ref{viscosity-soln}) \cite{wiener-1930}:
\begin{equation}
     \frac{\partial\phi}{\partial t}
      = -D\left(\frac{\partial\phi}{\partial x}\right)^2+i\epsilon D\left(\frac{\partial^2\phi}{\partial x^2}
      \right).
\end{equation}
Again let $\phi(x,t)=-i\epsilon\log\rho(x,t)$,
\begin{equation}
   i\frac{\partial\rho(x,t)}{\partial t} =
   -\epsilon D\frac{\partial^2\rho}{\partial x^2}.
\label{equation17}
\end{equation}
If we identify $\epsilon=\hbar$ and $D=1/2m$, then
this is Schr\"{o}dinger's equation of motion.  Eq. (\ref{equation17}) has eigenfunction $\rho_{\mu}(x,t)=e^{-i\epsilon D\mu^2 t+i\mu x}$ with eigenvalue $\epsilon D\mu^2$. Then
\[
   \frac{\partial}{\partial t} \big|\rho_{\mu}(x,t)\big|^2
   = \frac{\partial}{\partial t}\big( 1
   \big) = 0,
\]
for each and every $x$. This is interpreted as quantum {\em energy conservation}, and $|\rho(x,t)|^2$ is identified as probability density function.

The different choices of introducing the $\epsilon$ perturbation break Imaginary Scale symmetry that is inherent to the un-perturbed HJE.  They give rise to two different dynamic scenarios characterized by Eqs. (\ref{equation15}) and (\ref{equation17}) respectively.  No longer possessing the imaginary scale symmetry themselves, nevertheless, they are related via a Wick-rotation like operation.

The discussion of oscillatory wave solution with the imaginary ``$i$'' is not a mathematical fascination.  It is well known that there are two very different approaches to solve an HJE: (i) ``geometric point based" Hamiltonian trajectories as characteristic curves that require initial positions and momenta, and (ii) ``wave based'' solutions which require non-Dirac-$\delta$ initial $\phi(x,0)$ with finite support.  The latter has been the mathematical description of the wave front in optics.

\section{Hamilton's principle}

Realizing that $E=H(y)$ in Eq. (\ref{eq11}) is
incarnation of an HJE $\partial\phi/\partial t = -H(\partial\phi/\partial x)$ and the equality in Eq. (\ref{eq10}) holds true if and only if when $\psi(y,E)$ is the LF dual to $\phi(x,t)$, it is immediately clear that the former is the mathematical consequence of the latter via a variational principle as in thermodynamics.  This is analogous to the Hamilton's principle.

For a motion in $\mathbb{R}^{K}$, under the data {\em ad infinitum} as a fundamental postulate, there will be an entropy function $-\phi(\vx,t)$ where $\vx=(x_1,\cdots,x_K)$.  Then
\begin{eqnarray}
     P_{\vx}(t) &\propto& e^{-\phi(t,\vx)/\epsilon} =
     \exp\left(-\frac{t\phi(1,\vx/t)}{\epsilon} \right)
\nonumber\\
    &=& \exp\left(-\frac{1}{\epsilon}
    \sup_{E,\vy}\Big\{-tE+\vx\cdot\vy
    \Big| E= H(\vy) \Big\} \right)
\nonumber\\
    &=& \exp \left( -\frac{t}{\epsilon}
    \sup_{\vy}\left\{-H(\vy)+ \left(\frac{\vx}{t}\right)\cdot\vy\right\} \right)
\nonumber\\
    &=& \exp \left( -\frac{t}{\epsilon}
    L\big(\dot{\vx}\big) \right),
\label{eq18}
\end{eqnarray}
where $L(\dot{\vx})$ is a Langrangian and $\dot{\vx}=\vx/t$.  Generalizing (\ref{eq18}) to space without translational symmetry, a Lagrangian $L(\dot{\vx},\vx)$ has an explicit dependence on $\vx$, then Eq. (\ref{eq18}) becomes
\begin{equation}
    \exp\left(-\frac{1}{\epsilon}
    \int_0^t L\big(\dot{\vx}(\tau),\vx(\tau)\big) \rd \tau \right).
\end{equation}
The principle of least action then follows the principle of maximum entropy.

\section{Continuous-time limit and logarithmic Markov $\mP$  }

There is a deeper mathematical reason for Imaginary Scale symmetry breaking; it hints a ``topological divide'' that separates dissipative from conservative dynamics. Cyclic stochastic motions that can be mathematically represented as ``phase dynamics'' on a torus and captured by an imaginary $i$ signifies a different topology as dynamics on $\mathbb{R}^n$ \cite{qw_cmp}. A discrete state, discrete-time Markov process with transition matrix $\mP$ is consistent with dissipative Markov motions in continuous time if and only if the logarithm of the matrix $\mQ=\ln\mP$ is a proper Markov generator for continuous-time Markov chain: $\mP=e^{\mQ}$.  $\mQ$ is formally defined via the diagonalization of $\mP=\mB\MLambda\mB^{-1}$ where $\MLambda$ is a diagonal matrix and $\ln\mP=\mB(\ln\MLambda)\mB^{-1}$.  Therefore, only a $\mP$ whose Markovian $\mQ$ exists has a corresponding dissipative dynamics in continuous time.  This is an intrinsic characteristic of a $\mP$; it is revealed only through a full spectral analysis \cite{gardner-book}. This is the profound insight in J. B. J. Fourier's theory of heat equation\cite{fourier-book}\footnote{The spectral analysis for diffusion with drift yields $D\mu^2+V\mu=-\lambda$, where $\mu^{-1}$ and $\lambda^{-1}$ are the characteristic length and time in $e^{-\mu x-\lambda t}$. With non-dimensionalized $\tilde{\mu}=2D\mu/V$ and $\tilde{\lambda}=4D\lambda/V^2$, $\big(\tilde{\mu}+1\big)^2=-\tilde{\lambda}+1$, complex $\tilde{\mu}$ corresponds to $\tilde{\lambda}>1$.  On a finite interval $[0,L]$ with periodic boundary condition, $e^{i(2\pi k/L)x-\lambda t}$ has $D(2\pi k/L)^2+iV(2\pi k/L)=\lambda$, or $(k/k^*)^2+2i(k/k^*)=\tilde{\lambda}$, with a critical wave number $k^*=VL/(4\pi D)$ that divides dissipative from oscillatory motions. }; it is the justification for introducing the oscillatory wave solution in Sec. \ref{i-perturb}:  When $\mQ$ is complex, the corresponding dynamics in continuous time is not Markovian but rather has an oscillatory component with coherence in time.

A circulant $\mP$ matrix corresponds to translational symmetry with independent and identically distributed (i.i.d.) {\em increments}. It has all its eigenvalues being real if and only if the Markov dynamics is reversible. For irreversible $\mP$, $\mQ=\ln\mP$ in continuous time exhibits a ``fundamental frequency'' which agrees with translational symmetry in a continuous space $[0,L]$ with periodic boundary condition,
\begin{equation}
       \lambda_k=-D\left(\frac{2k\pi}{L}
        \right)^2-iV\left(\frac{2k \pi}{L}\right),
\end{equation}
where $k$ denotes an integer.  The suggested divide between dissipative and conservative motions can be further
studied in connection to the asymptotic integrals below, in the limit of $\epsilon\to 0$ \cite{bender-orszag-book}.  With smooth real functions $f(x)$ and $h(x)\ge 0$:
\begin{equation}
\label{eq21}
      \int_a^b f(x)e^{-h(x)/\epsilon} \rd x \sim f(x^*)e^{-h(x^*)/\epsilon} \left(\frac{2\pi\epsilon}{h''(x^*)}\right)^{1/2},
\end{equation}
in which $x^*=\arg\inf_{x\in\mathbb{R}}h(x)$ is assumed to be interior of $[a,b]$.  If $x^*$ is at $a$, the last factor on the rhs will be $\epsilon/h'(a)$.  In contrast,
\begin{equation}
\label{eq22}
      \int_a^b f(x)e^{i h(x)/\epsilon} \rd x \sim f(x^*)e^{ih(x^*)/\epsilon\pm i\pi/4}\left(\frac{2\pi\epsilon}{|h''(x^*)|}\right)^{1/2},
\end{equation}
where $h'(x^*)=0$ is the only stationary point in $[a,b]$. One uses $e^{i\pi/4}$ if $h''(x^*)>0$ and $e^{-i\pi/4}$ if $h''(x^*)<0$.  The asymptotic orders in Eqs. (\ref{eq21}) and (\ref{eq22}) are manifestly different, with $\mathcal{O}\big(\sqrt{\epsilon}\, e^{-\alpha/\epsilon}\big)$ where $\alpha>0$ and $\mathcal{O}\big(\sqrt{\epsilon}\big)$ respectively. They are so different that one only observes the latter with $\alpha=0$ in the former:  Sustained oscillations on the ``flat'' region of a landscape $\phi^{ss}(\vx)$ \cite{Ge-Qian-2012-Chaos,cheng-qian-jsp,qian_pla}.

\section{Discussion}

One of the legacies of P. A. M. Dirac is that one should take certain peculiar phenomena from a fundamental equation seriously.  We argue that Eq. (\ref{gHJE}), which has two parameters $p$ and $q$ if expressed in dimensionless homogeneous space and time $\tilde{x}=x/\eta$ and $\tilde{t}=\gamma t$, is fundamental.  The $\phi$ itself is an emergent quantity that is non-random in the continuous space and time limit. All the peculiar features associated with $\phi(x,t)$, as strong or weak solutions to (\ref{gHJE}), enrich our understanding of behaviors of the {\em entropy evolution}.  Taking Eq. (\ref{gHJE}) seriously was also the advice given in \cite{anderson72} on emergent phenomenon: ``Starting with the fundamental laws and a computer, we would have to do two impossible things | solve a problem with infinitely many bodies, and then apply the result to a finite system | before we synthesized this behavior.''  There are vast literature on HJE in both mathematics and theoretical physics; the present work provides a new vista from a thermodynamic perspective.

What is information?  How to quantify it?  Whatever it is, it grows exponentially with the ``extent'' of a system, which is the abstract $\epsilon^{-1}$ in the present work.  A system's physical size is universally used as its extent in classical thermodynamics \cite{callen-book}; for thermodynamics of small systems it was suggested that time is the essence for the large extent \cite{lu-qian-22}.  In mathematics, the deep relation between the Lyapunov exponent and entropy in nonlinear dynamics has been extensively studies \cite{qxz-book}.  In the age of machine learning, one identifies the extent with the amount of possible data. This work studies mechanical motions in which both time and space grow with information, the imagined data; they are extensive quantities.  Entropy $-\phi(x,t)$ is simply the growth rate of information per ``extent''.

In terms of kinematics vs. dynamics in classical mechanics, displacement $\Delta x$ and velocity $\dot{x}$ are concepts in the former while the notions of energy and momentum belong to the latter. Under the supposition that Langevin type stochastic movements are fundamental \cite{parisi-wu}, $V$ and $D$ are natural kinematic parameters in the limit of continuous space and time.  It follows that $\frac{1}{4}(\dot{x}-V)D^{-1}(\dot{x}-V)$, as the large deviation rate function w.r.t. $t$, suggests the notion of kinetic energy in dynamics, measured in temperature units $k_BT$ per unit time $\sim 2Dm$: Random motion is a manifestation of heat.  Furthermore, its Legendre-Fenchel transform is $Dy^2$, where $y$ is the conjugate variable to $\dot{x}$, with one-to-one correspondence $y=(\dot{x}-V)/(2D)$. It is tantalizing to fancy a deeper logic connection between the $D$ and ``mass'', the Newtonian dynamic notion that has been so fundamental that we cannot think of any other alternative interpretation.  Still, when matrix $D^{-1}(x)$ and vector $V(x)$ are multi-dimensional and being spatially dependent
without translational symmetry, one derives an equation of motion as
\begin{eqnarray*}
 \frac{\rd}{\rd t}\Big( D^{-1}_{ik}(x)\big(\dot{x}_k-V_k(x) \big)\Big) = \frac{1}{2}
 \big(\dot{x}_j-V_j(x)\big)\frac{\partial D^{-1}_{jk}(x)}{\partial x_i}
 \\
 \times\big(\dot{x}_k-V_k(x)\big) -\big(\dot{x}_j-V_j(x)\big)D^{-1}_{jk}(x) \frac{\partial V_k(x)}{\partial x_i},
\end{eqnarray*}
in which Einstein's summation rule is used.  A gauge-field theoretic analysis is forthcoming \cite{yangqianjsp}.

To conclude, the consequence of the present theory is manifold. We suggest one, possibly a more coherent modern mathematical narrative of microcanonical ensemble,  which plays a central role in the current teaching on the mechanical foundation of Gibbsian statistical thermodynamics.  We are motivated by the opening statement of \cite{pauli-v3}:
\begin{quotation}
{\em Classical thermodynamics foregoes detailed pictures ... Experiment shows that if a system is closed, then heat is exchanged within the system until a stable thermal state is reached.}
\end{quotation}
We now recognize that a given macroscopic closed mechanical system has two mathematical representations:

(i) a Hamiltonian representation based on Newtonian smooth motion captured by a detailed equation of motion that exhibits total mechanical (kinetic $+$ potential) energy $H$ conservation; and

(ii) a Gibbsian representation based on Brownian motion that is under the {\em conditioning} of total mechanical energy $H=E$; it foregoes detailed equation of motion. A proper conditional probability treatment requires an infinitesimal interval $[E,E+\rd E]$.

In the scenario (i) a linear Liouville equation is derived that describes a distribution $P_{\vx}(t)$ with deterministic motion in both continuous space and time.  However in sharp contrast, in the scenario (ii) a nonlinear HJE is derived, that describes the evolution of the emergent entropy function $-\phi(\vx,t)$.  The relationship between (i) and (ii) is the classical mathematical result that if an initial $\phi(\vx,0)$ is a Dirac-$\delta$ point mass $\delta\big(\vx-\hat{\vx}(0)\big)$, and when an additional initial $\dot{\vx}(0)$ is provided, then the solution to the HJE agrees with an $\vx(t)$ that follows the Hamiltonian dynamics in (i) \cite{goldstein-book}.

\section{Acknowledgement}
H. Q. thanks mathematical coworkers, Y. Gao (Purdue), L. Hong (SYSU), J. Hu (UW), J.-G. Liu (Duke), W. Sun (SFU) for extensive discussions.

\input{reference.bbl}
\bibliography{reference}

\appendix

\section{Gibbs-Hill's thermodynamics for macro- and nano-systems}
\label{app-A}

We give an expanded version for the very condensed three equations in Eq. (\ref{equation1}).  We only discuss systems with three thermodynamic variables and assume each system, macro- or nano-, has a concave entropy function $S(U,V,N)$ that is not necessarily Eulerian homogeneous.  Phase transition phenomenon is a separate topic that arises in systems with non-concave $S(U,V,N)$ \cite{qatw}.  If the $S$ is further assumed as an increasing function of $U$, as in most macroscopic thermodynamics, then $U(S,V,N)$ exists and it is a convex function of $S$, $V$, and $N$.

The subdivision potential function introduced by Hill in 1964 is the Legendre-Fenchel transform (LFT) \cite{Hill_SST_Book,lu-qian-22}:
\begin{eqnarray}
 \mathscr{E}(T,P,\mu) &=& T\inf_{U,V,N}\left\{\frac{U+PV-\mu N}{T}-S(U,V,N)\right\}
\nonumber\\
  &=& \inf_{N}\Big\{ \inf_{S,V}
  \Big\{ U+PV-TS(U,V,N) \Big\} -\mu N\Big\}
\nonumber\\
  &=& \inf_{N}\Big\{ G(T,P,N)-\mu N\Big\}.
\end{eqnarray}
Gibbs' $G(T,P,N)$ is a partial LFT of $S(U,V,N)$. Hill's $\mathscr{E}(T,P,\mu)$ is the full LFT of $S(U,V,N)$, and it is the LFT of $G(T,P,N)$ w.r.t. $N$ which is the {\bf\em last extensive variable}, a notion emphasized in \cite{lu-qian-22}.

For a macroscopic system whose $S(U,V,N)$ is an Eulerian homogeneous degree-$1$ function, $G(T,P,N)$ is a homogeneous function of $N$ alone, thus, it is strictly proportional to $N$: $G(T,P,N)=Nf(T,P)$, where
\begin{equation}
\label{23}
  \mu=f(T,P) = \left(\frac{\partial G(T,P,N)}{\partial N}\right)_{T,P} = \frac{G(T,P,N)}{N}.
\end{equation}
This is Gibbs' chemical potential.  Then
\begin{eqnarray}
  \mathscr{E}(T,P,\mu)=\inf_{N}\Big\{ Nf(T,P) -\mu N\Big\}
\nonumber\\
  = \left\{\begin{array}{ccc}
              0 &&  \mu=f(T,P)  \\
        -\infty &&  \mu\neq f(T,P)
        \end{array}\right.
\end{eqnarray}
which is known as the {\em indicator function} for $\mu=f(T,p)$, the equation of state (EoS).  The bivariate $\mu=f(T,P)$ is {\bf\em holographic} with respect to the trivariate $S(U,V,N)$; this follows the LF duality \cite{rockafellar-book} between homogeneous $S(U,V,N)$ and its corresponding $\mathscr{E}(T,P,\mu)$ whose support is singular.  Nevertheless one can obtain the former from the latter also by LFT:
\begin{widetext}
\begin{eqnarray}
  &&\inf_{T,P,\,\mu}\left\{\frac{U+PV-\mu N}{T}-\frac{\mathscr{E}(T,P,\mu)}{T}\right\} = \inf_{T,P}\left\{\frac{U+PV-Nf(T,P)}{T}\right\}
\nonumber\\
  &=& \inf_{T,P}\left\{ \frac{U+PV}{T}-\frac{1}{T}\inf_{U',V'}\Big\{ U'-TS(U',V',N)+PV'\Big\} \right\}
\nonumber\\
    &=& \sup_{U',V'}\left\{ S(U',V',N) +\inf_{T,P}\left\{\frac{U-U'}{T}+\frac{P(V-V')}{T}\right\} \right\}
\nonumber\\[4pt]
    &=& \sup_{U',V'}\left\{S(U',V',N) + \left\{\begin{array}{ccc}
        -\infty &&  U'\neq U \text{ or }
         V'\neq V \\
        0  && U'=U \text{ and }  V'=V
      \end{array}\right\} \right\}   \ = \ S(U,V,N).
\label{lftduality}
\end{eqnarray}
\end{widetext}

For convex (or concave) functions, because LFT yields a one-to-one, bijective relation between variables and their conjugate variables, chemical potential $\mu$ also has many other expressions as results from variable substitution:
\begin{equation}
    \mu = \left(\frac{\partial F(T,V,N)}{\partial N}\right)_{T,V}
     =f\big(T,P(T,V,N)\big),
\end{equation}
where
\begin{eqnarray}
    &\displaystyle
     F(T,V,N) = \inf_{P}\Big\{G(T,P,N)-PV\Big\},
\\
    &\displaystyle
     P(T,V,N) = -\frac{\partial F(T,V,N)}{\partial V}.
\label{equation26}\\
    &\displaystyle
     G(T,P,N) = \sup_{V}\Big\{ F(T,V,N)+PV\Big\},
\\
     &\displaystyle
     V(T,P,N) = \frac{\partial G(T,P,N) }{\partial P},
\label{equation25}
\end{eqnarray}
The duality between $F(T,V,N)$ and $G(T,P,N)$ can be verified following the same mathematics as in (\ref{lftduality}) by substituting (\ref{equation25}) into (\ref{equation26}).  All the other functional expressions for $\mu$ are for the same reason:
\begin{eqnarray}
   \mu &=& \left(\frac{\partial H(S,P,N)}{\partial N}\right)_{S,P} =f\big(T(S,P,N),P\big),
\\
  \mu &=& \left(\frac{\partial U(S,V,N)}{\partial N}\right)_{S,V} = f\Big(T(S,V,N),P(S,V,N)\Big).
\nonumber\\
\end{eqnarray}

This kind of results of course is not unique for $\mu$.  Instead of $N$ in Eq. \ref{23} being the last extensive variables, if one considers $U$ as the last extensive variable, then temperature $T$ first appears in the duality between entropy and Helmholtz free energy of isothermal ensemble:
\begin{eqnarray}
      F(T,V,N) &=& T\inf_{U}\left\{ \frac{U}{T}-S(U,V,N)\right\},
\\
    U(T,V,N) &=& \frac{\partial(F(T,V,N)/T)}{\partial (1/T)},
\\[6pt]
      S(U,V,N) &=& \inf_{T}\left\{ \frac{U}{T}-\frac{F(T,V,N)}{T}\right\},
\\
  T^{-1}(U,V,N) &=& \frac{\partial S(U,V,N)}{\partial U}.
\end{eqnarray}
One can also have isobaric grand ensemble
\begin{eqnarray}
  \Xi(U,\tilde{P},\tilde{\mu}) &=& \sup_{V,N}\Big\{S(U,V,N) -\tilde{P}V+\tilde{\mu}N\Big\}
\\
  &=& U\left(\frac{\partial \Xi(U,\tilde{P},\tilde{\mu})}{\partial U}\right)_{P,\mu} = U\beta(\tilde{P},\tilde{\mu}).
\nonumber
\end{eqnarray}
That is,
\begin{eqnarray}
    \frac{1}{T}= \beta(\tilde{P},\tilde{\mu}) &=& \left(\frac{\partial\Xi(U,\tilde{P},\tilde{\mu})}{\partial U}\right)_{\tilde{P},\tilde{\mu}}
\nonumber\\
    &=& \frac{\Xi(U,\tilde{P}, \tilde{\mu})}{U}.
\end{eqnarray}
Actually the relation $T=\beta^{-1}(P/T,\mu/T)$ is simply a variation of the $\mu=f(T,P)$ in Eq. \ref{23}.  One can also express $T$ in many other functional forms.  They could all be used as exercises for a college class on thermodynamics.

\section{A simple Langevin dynamics under a constraint}
\label{app-B}

Newtonian motion is a smooth trajectory of a geometric point. In the light of modern understanding of H. Poincar\'{e}'s nonlinear dynamics with chaotic motions, this perspective is unattainable.  Motions in continuous space and time need to be represented by an entire distribution starting from the neighborhood $\delta_x$ of an initial point $x$ that takes the ideas from modern mathematics of measure theory and weak solutions to a PDE into account; this is the significance of a Liouville
equation and the notion of ergodic dynamics \cite{quay_ergodicity}: There are now two limits $\delta_x\to x$ and $t\to\infty$ in the description; taking $\delta_x\to x$ first yields the Newtonian motion but loses the thermodynamic description; taking $t\to\infty$ first can lead to an invariant Gibbs measure \cite{ruelle-book,qxz-book}.  An HJE in terms of neg-entropy $\phi(x,t)$ represents the motion on an exponential scale without randomness.  The exponential nature arises from the limit of $\epsilon\to 0$ (or $\delta_x$) conditioned on a deterministic ``zero probability'' event; this is the fundamental spirit of large deviation theory \cite{dembo-book}.

To wit, let us consider a simple stochastic motion in terms of a Langevin equation with constant drift $V$ and diffusion coefficient $\epsilon D$: $\rd X(t) = V\rd t + \sqrt{2\epsilon D}\,\rd W(t)$, where $W(t)$ is the Brownian motion and $\xi(t)=\rd W/\rd t$ is a white noise with $\langle \xi(t)\xi(t+\tau) \rangle=\delta(\tau)$. We focus on $X(t)$ with $t\in [0,T]$; in the modern theory of probability one considers the set that contains all the continuous functions $\Omega=\{X(t): 0\le t\le T\}$, and the probability of the entire path is \cite{ge_qian_ijmpb}
\begin{eqnarray}
\label{B1}
   && \mathbb{P}\big\{ X(t) = x(t), 0\le t\le T \big\}
\\
   &=& \mathcal{A}(\epsilon)
   \exp\left\{-\frac{1}{4\epsilon D} \int_0^T\Big(\dot{x}(s)-V\Big)^2\rd s\right\},
\nonumber
\end{eqnarray}
in which $x(\cdot)\in\Omega$ and $\mathcal{A}(\epsilon)$ is a normalization factor.

Among all the $X(t)$ with $X(0)=x_0$, the most probable path is $X(t)=Vt+x_0$, which maximizes (\ref{B1}). Now let us consider all the $X(t)$ with $X(0)=x_0$ and $X(T)=x_1\neq x_0+VT$, and then take the limit $\epsilon\to 0$.  The resulting $X(t)$ is a non-random function which maximizes the conditional probability, {\em i.e.}, minimizes the action functional,
\begin{equation}
   S\big[x(t), 0\le t\le T\big] =
   \int_0^{T} \Big( \dot{x}(s)-V\Big)^2
     \rd s,
\end{equation}
under the two constraints.  If $x(t)$ is twice differentiable then following Euler-Lagrange equation, one has
\begin{equation}
   \frac{\rd}{\rd t} \dot{x}(t) =
   \frac{\rd^2 x(t)}{\rd t^2} = 0,
\end{equation}
whose solution now is $x(t)=x_0+(x_1-x_0 )t/T$.  This, we suggest, is a rather novel interpretation of Newton's law of inertia as an emergent phenomenon {\em \`{a} la} Anderson \cite{anderson72}.  The HJE in the main text is the result of a systematic treatment of this simple example; it initiates an alternative understanding of the relationship between HJE and Newton's equation of motion \cite{ge_qian_ijmpb}.

\section{Derivation of Entropy Evolution Equation (\ref{gHJE})}
\label{app-C}

In the continuous space and time limit of $\Delta x$, $\Delta t\to 0$, large deviation theory (LDT) \cite{dembo-book} asserts that $P_x(t) \sim e^{-\phi(x,t)/\epsilon}$. Substituting this form into the time stepping of the probability distribution leads to
\begin{eqnarray}
 e^{-\phi(x,t+\Delta t)/\epsilon} &=&
 p e^{-\phi(x-\Delta x,t)/\epsilon} + re^{-\phi(x,t)/\epsilon}
 \nonumber\\
  &+& q e^{-\phi(x+\Delta x,t)/\epsilon},
\end{eqnarray}
in which $r=1-p-q$.  Then
\begin{eqnarray}
e^{-[\phi(x,t+\Delta t)-\phi(x,t)] /\epsilon} &=& p e^{-[\phi(x-\Delta x,t)-\phi(x,t)]/\epsilon} + r
\nonumber\\
&+& q e^{-[\phi(x+\Delta x,t)-\phi(x,t)]/\epsilon}.
\label{eqb2}
\end{eqnarray}
In the limit of $\Delta t$, $\Delta x$, and $\epsilon\to 0$ all in the same order with $\gamma = \epsilon/\Delta t$ and $\eta=\Delta x/\epsilon$ fixed, we have a limit of (\ref{eqb2})
\begin{equation}
e^{-\gamma^{-1}\frac{\partial \phi}{\partial t}} = pe^{\eta \frac{\partial\phi}{\partial x} } + r +qe^{-\eta\frac{\partial\phi}{\partial x}},
\end{equation}
rearranging of which results in (\ref{gHJE}).

We also give an alternative derivation of Eq. (\ref{gHJE}); it provides a geometric optics perspective. In a Markov dynamics $X(t)$, conditional increments are considered i.i.d. random events with probability $\mathbb{P}\big\{ \Delta X \in \rd\xi \,|\, X(t)=x \big\}$ with $\Delta X \equiv X(t+\Delta t)-X(t)$.  If the state ``$X=x$'' is repeated sampled with a large number of realizations ($N=\epsilon^{-1}$ with $\gamma=1$ as time unit) from a recurrent, ergodic dynamics \cite{quay_ergodicity}, the Law of Large Numbers (LLN) defines the mean velocity at $x$, $V(x)=(\Delta t)^{-1} \mathbb{E}^x\big[\Delta X(t) \big]$, and the Central Limit Theorem (CLT) represents its Gaussian fluctuations
$D(x)=(2\Delta t)^{-1} \mathbb{E}^x\big[\big(\Delta X-V(x)\Delta t\big)^2\big]$, where $\mathbb{E}^x[\cdots]$ is the conditional expectation.  What is not widely appreciated is that LLN and CLT are partial results of the much information richer mathematical characterization of $\Delta X$: LDT states that in the same limit of LLN and CLT as the sample size $N\to\infty$,
\begin{equation}
    \mathbb{P}\big\{ \Delta X(t) \in \rd\xi  \,\big| X(t) = x  \big\} \simeq e^{-NI(\xi;x)}\rd \xi,
\label{L-function}
\end{equation}
where $I(\xi;x)\ge 0$ has its global minimum $I(V(x)\Delta t;x)=0$ with curvature $\big(2D(x)\big)^{-1}$.  Introducing entropy function $\phi(x,t)=-N^{-1}\log\mathbb{P}\{X(t)\in\rd x\}$, the law of total probability $\mathbb{P}\big\{X(t+\Delta t)\in\rd x \big\}=$
\[
   \int_{\mathcal{M}}
    \mathbb{P}\big\{X(t+\Delta t) \in \rd x \,\Big|\, X(t)=x'\big\}
     \mathbb{P}\big\{X(t) \in \rd x' \big\},
\]
in the asymptotic limit yields
\begin{equation}
    \phi(x,t+\Delta t) = \inf_{x'}
    \Big\{ I\big(x-x' ;x'\big)
      + \phi(x',t)  \Big\}.
\end{equation}
This relation is known as Bellman's
Principle of Optimality in which $\phi$ is called the cost-to-go function. When the $\Delta t$ and $x-x'$ are infinitesimal and noting that for i.i.d. increments $I(x-x';x')=I(x-x')$, one has
\begin{eqnarray}
 \frac{\partial\phi}{\partial t}  &=& -\sup_{\dot{x} }
    \left\{ \dot{x}\, \frac{\partial\phi}{\partial x}(x,t) - \frac{1}{\Delta t} I\big( \dot{x}\Delta t\big)
       \right\}
\nonumber\\
  &=& -\sup_{\dot{x}}\big\{ \dot{x}y-L\big(\dot{x}\big)\big\}
 = -H(y),
\label{eqb6}
\end{eqnarray}
in which $\dot{x}=(x-x')/\Delta t$, $y\equiv\partial\phi/\partial x$,
and $I(\dot{x}\Delta t)\to \Delta t\, L(\dot{x})$. $H(y)$ is the Legendre-Fenchel transform of $L(\dot{x})$, the cumulant generating function of $\dot{x}=0,\pm\eta$:
\begin{equation}
\label{Hwithoutx}
  H(y)=\log\Big(pe^{\eta y}+r+qe^{-\eta y}\Big).
\end{equation}
Eq. (\ref{eqb6}) is the HJE in (\ref{gHJE}), with $\gamma=1$.

\section{The physics of continuous space and time limit}

As $\Delta x$, $\Delta t \to 0$, Newtonian smooth motion simply assumes the existence of
$\Delta x/\Delta t = v$.  Stochastic motion has increment in $\Delta t$ time, $X(t+\Delta t)-X(t)=\Delta x$, $-\Delta x$, $0$ with probabilities $p$, $q$, $r$, and $p+q+r=1$.  The
canonical diffusion limit supposes the existence of
\begin{eqnarray}
    &\displaystyle
   \frac{(p-q)\Delta x}{\Delta t} = V, \ \frac{(1-r)(\Delta x)^2}{2\Delta t} = D, \text{ and}
\nonumber\\[-6pt]
\\[-6pt]
    &\displaystyle
    \frac{1}{\Delta x}
    \mathbb{P}\big\{X(t)\in
    [\,x,x+\Delta x\,]\,\big\} = f(x,t),
\nonumber
\end{eqnarray}
as $\Delta t\to 0$. This imply $\frac{ 2D(p-q)^2  }{(1-r)V^2 } \sim \Delta t$.  If, however, $p,q,r$ are independent of $\Delta t$, then the existence of $D$ implies
$\frac{\Delta x}{\Delta t}\to\infty$. The resulting {\em overdamped} Brownian motion $B(t)$ is nowhere differentiable \cite{DurrettBook}.

In contradistinction the present work supposes the existence of
\begin{eqnarray}
    &\displaystyle
  \frac{\Delta x}{\epsilon} = \eta, \  \frac{\Delta t}{\epsilon} = \gamma^{-1}, \text{ and}
\nonumber\\[-6pt]
\\[-6pt]
    &\displaystyle
  -\epsilon \log \mathbb{P}\big\{X_{\epsilon}(t)\in \rd x\big\} = \phi(x,t),
\nonumber
\end{eqnarray}
as $\epsilon\to 0$; thus $\Delta x/\Delta t=\gamma\eta$ exists.  We are interested in the motion on $\epsilon^{-1}$ scale; the $\phi$ satisfies an HJE with the Hamiltonian function given in Eq. (\ref{Hwithoutx}).  The corresponding Hamiltonian dynamics then is
\begin{equation}
\label{xdot}
  \dot{x} = \frac{\gamma\eta\big( pe^{\eta y} - q e^{-\eta y}\big) }{ pe^{\eta y} + r + q e^{-\eta y} }, \ \
  	  \dot{y} = 0.
\end{equation}
$\dot{x}$ given above is a monotonic function of $y$ and $|\dot{x}|\to \gamma\eta$ as $|y|\to\infty$.  One now has $|\dot{x}|\le \gamma\eta$, a limit that is independent from $p,q,r$.  This is the meaning of {\em undamped/conserved} Hamiltonian motion.

Introducing $\tilde{v}\equiv \dot{x}/(\gamma\eta)$, and solving $y$ in terms of $\tilde{v}$ from (\ref{xdot}), $|\tilde{v}|\le 1$:
\begin{equation}
 y = \frac{1}{\eta}\log\left\{\frac{r\tilde{v}+\sqrt{4pq-(4pq-r^2)\tilde{v}^2 } }{2p\big(1-\tilde{v}\big) }
    \right\}.
\end{equation}
For isotropic space with $p=q=(1-r)/2$, $y(\tilde{v})$ is an odd function.  As $\tilde{v}\to 0$, $y\simeq\tilde{v}/(\eta(1-r))$ and as $|\tilde{v}|\to 1$,
\begin{eqnarray}
  y &=& \frac{1}{2\eta}\log\left\{ \frac{1-\tilde{v}}{1+\tilde{v}}\, \frac{\sqrt{r^2+(1-2r)(1-\tilde{v}^2) }-r\tilde{v}}{\sqrt{r^2+(1-2r)(1-\tilde{v}^2) }+r\tilde{v} }
    \right\}
\nonumber\\
    &\simeq& \frac{1}{2\eta}\log\left\{ \left(\frac{1-\tilde{v}}{1+\tilde{v}}\right)^2\frac{2r^2+(1-2r)(1+\tilde{v}) }{2r^2+(1-2r)(1-\tilde{v}) } \right\}
\nonumber\\
    &\simeq& \frac{1}{\eta}\log \left(\frac{1-\tilde{v}}{1+\tilde{v}}\right)
\\
    &+& \frac{\tilde{v} }{\eta|\tilde{v}|}\left\{ \log\frac{1-r}{r}-a_1\big(1-|\tilde{v}|\big)+a_2\big(1-|\tilde{v}|\big)^2 \right\},
\nonumber
\end{eqnarray}
where $a_1=\frac{(1-2r)(1-2r+2r^2)}{4r^2(1-r)^2}$ and $a_2=\frac{(1-2r)^2[(1-r)^4-r^4]}{16r^4(1-r)^4}$.

\end{document}

%% file: macros.tex
\def\rd{{\rm d}}

\def\vp{{\bf p}}
\def\vq{{\bf q}}

\def\vx{{\bf x}}
\def\vy{{\bf y}}

\def\MLambda{\boldsymbol{\Lambda}}

\def\mB{{\bf B}}

\def\mP{{\bf P}}
\def\mQ{{\bf Q}}

%% file: reference.bbl
%